\documentclass{jltp}

\usepackage{graphicx} 

\title{Analytical solution for nonlinear Schr$\ddot{\hbox{o}}$dinger 
vortex reconnection}
\author{Sergey Nazarenko and Robert West}
\address{Mathematics Institute, University of Warwick, Coventry, CV4 7AL, UK}

\runninghead{S.V. Nazarenko and R.J. West}{Analytical solution for 
nonlinear Schr$\ddot{\hbox{o}}$dinger vortex reconnection}

\begin{document}

\maketitle


\begin{abstract}
Analysis of the nonlinear Schr$\ddot{\hbox{o}}$dinger vortex reconnection
is given in terms of coordinate-time power series. The lowest order terms
in these series correspond to a solution of the linear
Schr$\ddot{\hbox{o}}$dinger equation and provide several interesting properties
of the reconnection process, in particular the non-singular character of
reconnections, the anti-parallel configuration of vortex filaments and a
square-root law of approach just before/after reconnections. The complete
infinite power series represents a fully nonlinear analytic solution in a finite
volume which includes the reconnection point, and is valid for finite time provided
the initial condition is an analytic function. These series solutions are
free from the periodicity artifacts and discretization error of the direct
computational approaches and they are easy to analyze using a computer algebra
program.

PACS number: 67.40.Vs
\end{abstract}


\section{INTRODUCTION}

Vortex solutions of the nonlinear Schr$\ddot {\hbox{o}}$dinger 
(NLS) equation are of interest in nonlinear optics
\cite{{Akhmediev},{Snyder},{Swartzlander}}
and in the theory of Bose-Einstein condensates \cite{{Berloff}} (BEC).
The NLS equation is also often used to describe turbulence in superfluid
helium. \cite{gp} NLS is a nice model in this case 
because the vortex quantization appears naturally in this model
and  because
its large-scale limit is the compressible Euler equation 
describing classical inviscid fluids. \cite{{Schwarz},{Ercolani}}
At short scales, the NLS equation
allows for a ``quantum uncertainty principle'' which allows
vortex reconnections without the need for a finite 
viscosity or other dissipation. Numerically, NLS vortex
reconnection was studied by Koplik and Levine\cite{Koplik} and, more recently,
by Leadbeater {\em et al.}\cite{Leadbeater} and, for a non-local version of NLS 
equation, by Berloff {\em et al.}\cite{Berloff} In applications to superfluid 
turbulence, the NLS equation was directly computed by Nore {\em et al.}\cite{Nore} 
Such cryogenic turbulence consists of repeatedly reconnecting vortex 
tangles, with each reconnection event resulting in the generation of 
Kelvin waves on the vortex cores \cite{Svistunov} and a 
sound emission. \cite{Leadbeater}
These two small-scale processes are very hard to correctly compute
in direct simulations of 3D NLS turbulence due to 
numerical resolution problems. A popular way to avoid this
problem is to compute vortex tangles by a Biot-Savart method
(derived from the Euler equation) and use a simple rule to reconnect 
vortex filaments that are closer than some critical (``quantum'') distance. 
This approach was pioneered by Schwarz \cite{Schwarz} and it has been further developed by
Samuels {\em et al.}\cite{Barenghi} In this case, it is important
to prescribe  realistic  vortex reconnection rules.
Therefore, elementary vortex reconnection events have
to be carefully studied and parameterized. Numerically, such a
study was performed by Leadbeater {\em et al.}, \cite{Leadbeater} the present paper 
is devoted to the analytical study of these NLS vortex reconnection events.

The analytical approach of this paper is based on expanding 
a solution in powers of small distance from the
reconnection point, and  small time measured from the 
reconnection moment. 
The idea is to exploit the fact that when vortex filaments are near reconnection, 
the nonlinearity in the NLS equation is small.  This smallness of the nonlinearity 
just stems from the definition of vortices in NLS (curves where $\Psi=0$) and the 
continuity of $\Psi$. Their core size is of the order of the distance over 
which $\Psi\rightarrow 1$ (where $\Psi=1$ represents the background condensate). 
Therefore, for vortices near reconnection, separated by a distance much smaller than 
their core size, $\Psi$ is small provided it is continuous. 
%
%
Thus, to the first approximation the
solution near the reconnection point can be described 
by a linear solution which, already at this level, contains
some very important information about the reconnection process:
(1) that the reconnection proceeds smoothly without any singularity
formation, (2) that in the immediate vicinity of the 
reconnection the vortices are strictly anti-parallel and
(3) just before the reconnection event the distance between the vortices
decreases as $|t|^{1/2}$, where $t$ is the time measured from the
reconnection moment. Note that result (1) could surprise those who draw their
intuition from vortex collapsing events in the Euler equation (which are believed
to be singular). On the other hand, results (2) and (3) are remarkably
similar to the numerical and theoretical results found for the Euler equation. \cite{{Pumir1},{Pumir2},{Pumir3}}

In section II of this paper we examine the local analysis of the 
reconnection process by deriving a linear solution and in section III
consider its properties. The linear solution describes many, but not all the 
important properties of vortex reconnection. In particular, it cannot describe 
solutions outside the vortex cores and, therefore, it cannot
describe the far-field sound radiation produced by the reconnection. On the other hand,
one can substitute the linear solution back into the NLS equation and
find the first nonlinear correction to this solution. Recursively
repeating this procedure, one can recover the fully nonlinear
solution in terms of infinite coordinate and time series. This derivation is
discussed in detail in section IV. The series produced are a general solution to a
Cauchy initial value problem.
Thus, by Cauchy-Kowalevski theorem, \cite{Cauchy} these series define an
analytic function (with a finite convergence radius) provided
the initial conditions are analytic. The generation of such a suitable 
initial condition is addressed in section V. Our series representation 
of the solution to the NLS equation is exact, and therefore will
include such properties as sound emission. However,
due to the finite radius of convergence of the analytic solution,
one is unable to observe a far-field sound emission directly. 
In this paper, we use {\em Mathematica} to compute some examples of the fully nonlinear
solutions for the vortex reconnection. The results of which are presented in
section VI.

Let us summarise the advantages and disadvantages that our
analytical solution has with respect to those being 
computed via direct numerical simulations (DNS).
Firstly, our analytical solutions are obtained as a
general formula, simultaneously applicable for a broad
class of initial vortex positions and orientations.
Secondly, our analytical solutions are not affected by
any periodicity artifacts (which are typical in DNS using
spectral methods) or by discretization errors.
On the other hand, our analytical solutions are only
available for a finite distance from the vortex lines
(of the order of the vortex core size) because their
defining power series have a finite radius of convergence.

\section{LOCAL ANALYSIS OF THE RECONNECTION}

Let us start with the defocusing NLS equation written in the 
non-dimensional form,
\begin{equation}\label{NLS}
i\Psi_t+\Delta \Psi +(1-|\Psi|^2) \Psi =0.
\end{equation}
Suppose that in vicinity of the point ${\bf r}=(x,y,z)=(0,0,0)$
at $t=t_0$ we have $\Psi = \Psi_0$ such that
$Re \Psi_0 = z$, and $Im \Psi_0 = az+bx^2-cy^2$, 
%
where $a,b$ and $c$ are some positive constants. For such initial conditions
the geometrical location of the vortex filaments, $\Psi=0$,
is given by two intersecting straight lines,
$z=0$ and $y=\pm\sqrt{b/c}\, x$.
%

\begin{figure}[!Ht]
\centerline{\includegraphics[width=4.5in]{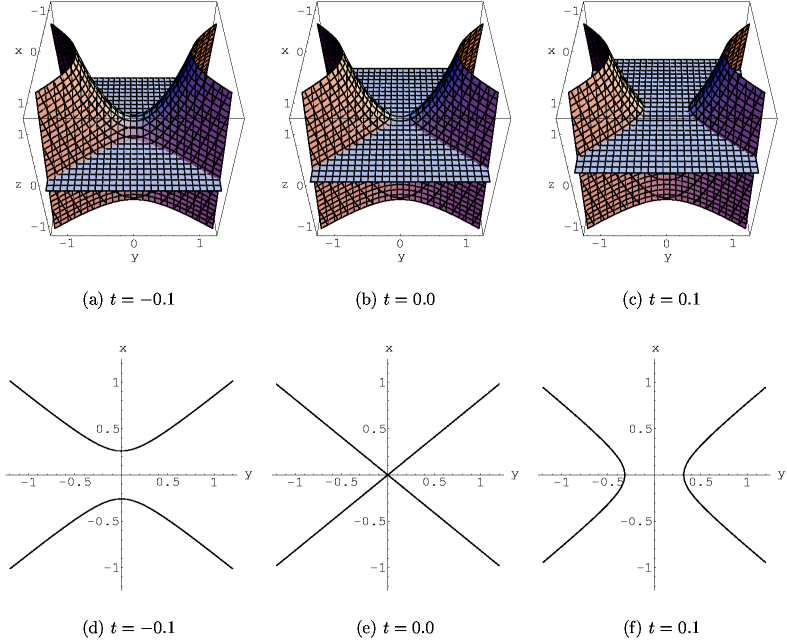}}
 \caption{Linear solution Eq. (\ref{linear}) of the nonlinear Schr$\ddot{\hbox{o}}$dinger 
          equation for  $a=1$, $b=3$, $c=2$ and $t_0=0$. Sub-figures (a), (b) and (c) show 
          the intersection of the real (plane) and imaginary (hyperbolic paraboloid) parts 
          of Eq. (\ref{linear}) at successive times \mbox{$t=-0.1$}, \mbox{$t=0.0$} and 
          \mbox{$t=0.1$} respectively. Sub-figures (d), (e) and (f) show the corresponding 
          lines of intersection where \mbox{$\Psi=0$}; reconnection occurs at 
          \mbox{$t=t_0=0$}.}
   \label{contour}
\end{figure}

In the small vicinity of the point ${\bf r} =0$, deep inside 
the vortex core (where $\Psi_0 \approx 0$), we can ignore
the nonlinear term found in equation (\ref{NLS}). Further, by a simple 
transformation $\Psi = \Phi e^{it}$ we can eliminate the 
third term $\Psi$ and obtain
$i\Phi_t+\Delta \Phi = 0$.
%
%
(This just corresponds to multiplying our solution by a phase, it does not alter 
its properties, but does make the following analysis simpler).
It is easy to see that the initial condition has not changed under this transformation,
$\Psi_{0} = \Phi_{0}$. Advancing our system a small distance in time $t-t_0$, we find
$Re\,\Phi = Re\,\Psi_0 - (t-t_0) \,\Delta Im\,\Psi_0$ and
$Im\,\Phi = Im\,\Psi_0 + (t-t_0) \,\Delta Re\,\Psi_0$,
%
%
or
\begin{eqnarray}\label{linear}
Re\,\Phi &=& z-2(b-c)\, (t-t_0),\\
Im\,\Phi &=& az+bx^2-cy^2. \nonumber
\end{eqnarray}
For both $t-t_0<0$ and $t-t_0>0$ the set of vortex lines, $\Phi=0$,
is given by two hyperbolas. A bifurcation happens at
$t=t_0$ where these hyperbolas degenerate into the two intersecting
lines (see Fig. \ref{contour}). This bifurcation corresponds to
the reconnection of the vortex filaments. Thus, we have constructed
a local (in space and time) NLS solution corresponding to
vortex reconnection. Obviously, this solution corresponds to a 
smooth function $\Phi$ at the reconnection point. It should be stressed that
this is not an assumption, but just the way in which we have chosen to 
construct our solution. However, we do believe that this observed
smoothness is a common feature of NLS vortex
reconnection events. If this is true then all such reconnecting
vortices could locally be described by the presented solution as
the intersection of a hyperbola with a moving plane provides a generic
local bifurcation describing a reconnection in the case of smooth fields.
%
%

\section{PROPERTIES OF THE VORTEX RECONNECTION}

The local linear solution we have constructed (\ref{linear}) reveals 
several important properties of the reconnection of NLS vortices.
\newline
1. Whatever the initial orientation of the vortex filaments,
the reconnecting parts of these filaments align so that they approach
each other in an anti-parallel configuration. Indeed, according to
(\ref{linear}), the fluid velocity field
$
   \vec{v} = \nabla \arctan ( [Im\,\Phi]/[Re\,\Phi] )
           = \nabla \arctan ( [az + bx^2 -cy^2]/[z-2(b-c)(t-t_0)] )
$.
At the mid-point between the two vortices one finds
a velocity field consistent with an anti-parallel pair, 
$
   \vec{v} = 1/[2a^2(c-b)(t-t_0)] \vec{e}_z \neq 0.
$
(For a parallel configuration one would find $\vec{v} = 0$).
Amazingly similar anti-parallel configurations
have been observed in the numerical Biot-Savart simulations of thin 
vortex filaments 
in inviscid incompressible fluids.\cite{{Pumir1},{Pumir3}} 
\newline
2.
The reconnecting parts of the
vortex filaments approach each other as $\sqrt{t-t_0}$. 
Indeed, setting $Re\,\Phi = Im\,\Phi = 0$ and $y=0$ in 
(\ref{linear}) one 
obtains
$
  x = \pm \sqrt{2a[(c/b)-1](t-t_0)}.
$
Exactly the same scaling behaviour, for approaching thin filaments in
incompressible fluids, has been given by the theory of Siggia and
Pumir \cite{{Pumir1},{Pumir2},{Pumir3}} and has been observed numerically 
in Biot-Savart computations. \cite{{Pumir1},{Pumir3}}
\newline
3. The nonlinearity plays a minor role in the late
stages of vortex reconnection in NLS.
This is a simple manifestation of the fact that in the
 close spatio-temporal
vicinity of the reconnection point $\Psi \approx 0$,
so that the dynamics are almost linear. 
\newline
This last property can be also reformulated as follows. No singularity is observed in the 
process of reconnection according to the solution (\ref{linear}): 
both the real and imaginary parts of $\Psi$ behave continuously 
in space and time. This property is in drastic contrast to the 
singularity formation found in vortex collapsing events described by 
the Euler equation. Indeed, distinct from incompressible fluids, 
no viscous dissipation is needed for the NLS vortices to reconnect. 
Here, dispersion does the same job of breaking the topological constraints (related to
Kelvin's circulation theorem) as viscosity does in a normal fluid.

\section{NONLINEAR SOLUTION}
We will now move on to consider the full NLS equation. We will use a recursion
relationship to compute the solution assuming that \mbox{$x,y,z\sim \epsilon$}
and \mbox{$t\sim \epsilon^{3}$} (for simplicity, we take $t_0=0$). 
The solution we obtain will therefore be of the form 
\mbox{$\Psi=\Psi^{(0)}+\Psi^{(1)}+\Psi^{(2)}+\cdots$}, where 
\mbox{$\Psi^{(n)}\sim\epsilon^{n}$}.
The above $\epsilon$ scaling of $x$, $y$, $z$ and $t$ has been chosen to generate
a recursion relationship when substituted in the NLS equation (\ref{NLS}). Of course
we could have chosen a different $\epsilon$ dependence, however, as the final series representation
of our solution contains an infinite number of terms, this would just correspond to
the same solution but with a suitable re-ordering.

Consider the NLS equation (\ref{NLS}). Firstly, we note that 
\mbox{$\partial_{t}\sim \epsilon^{-3}$} and
\mbox{$\triangle\sim\epsilon^{-2}$} and therefore
   $i\Psi_{t}^{(m)} \sim\epsilon^{m-3}$,
   $\triangle\Psi^{(n)} \sim\epsilon^{n-2}$,
   $\left[ |\Psi|^{2}\Psi \right]^{(p)} \sim\epsilon^{p}$ and
   $\Psi^{(q)} \sim\epsilon^{q}$,
%
%
where
%
$
 [|\Psi|^{2}\Psi]^{(p)}
 = \sum_{i,j=1}^{p}\Psi^{*(i)} \Psi^{(j)} \Psi^{(p-i-j)}
$.
%
Matching the terms, by setting $m=n+1$ and $p=q=n-2$, and integrating we find
\begin{equation}\label{t-rec}
 \Psi^{(n+1)}= \Psi_{0}^{(n+1)}
             + i\int_{0}^{t} \left[\triangle\Psi^{(n)}
             +\Psi^{(n-2)}-[|\Psi|^{2}\Psi]^{(n-2)}\right] dt, 
\end{equation}
where \mbox{$\Psi_{0}^{(n)}$} are arbitrary $n^{th}$ order functions of
coordinate which appear as constants of integration with respect to time.
The full nonlinear solution of the Cauchy initial value problem can now
be obtained by matching $\Psi_{0}^{(n)}$ to the $n^{th}$ order components
of the initial condition at $t=0$ obtained via a
Taylor expansion in coordinate. Let us assume that the initial condition
is an analytic function so that it can be represented by power series 
in coordinates with  a non-zero volume of convergence. Then, by the  
Cauchy-Kowalevski theorem, the function $\Psi$ will remain analytic for 
non-zero  time. In other words, the solution can also be represented 
as a power series with a non-zero domain of convergence in space and time.
Remarkably, the recursion relation Eq. (\ref{t-rec}) is precisely the 
means by which one can write down the terms of the power-series 
representation of the fully nonlinear solution to the NLS equation, with
an arbitrary analytical initial condition $\Psi_{0}$.

\section{INITIAL CONDITION}
Our next step is to construct a suitable initial condition for our study of 
reconnecting vortices. This initial condition will have to be formulated in 
terms 
of a power series. We start by formulating the famous line vortex 
solution to the steady state NLS equation \cite{gp} in terms of a power 
series. Substituting \mbox{$\Psi=Ae^{i\theta}$} into Eq. (\ref{NLS}), we 
find
$   \triangle A - A|\nabla\theta|^{2} + A - A^{2}=0\nonumber$,
%
%
where we have used the fact that \mbox{$\nabla\theta\cdot\nabla A = 0$} and 
\mbox{$\triangle \theta=0$}. We can simplify this equation, since $A=A(r)$ 
and therefore, \mbox{$\triangle A = \frac{1}{r}\partial_{r}(r\partial_r A)$}. 
However, we also note that \mbox{$|\nabla^{2}\theta |=\frac{1}{r^{2}}$} since 
\mbox{$\partial_x \theta =-y/r^{2}$} and \mbox{$\partial_y \theta =-x/r^{2}$}. 
Therefore, we have
   $\frac{1}{r}\partial_{r}(r\partial_r A)-\frac{A}{r^{2}}-A^{3}+A=0$.
%
%
We will solve this equation using another recursive
method. We would like to get a solution of the form 
$A= a_{0} + a_{1}r + a_{2}r^{2} + a_{3}r^{3} + \cdots = \sum_{n} A^{(n)}$.
(However, we can set $a_{0}$ to zero on physical grounds, since we require 
$\Psi=0$ at $r=0$). As before
  $\frac{1}{r}\partial_{r}(rA_{r}^{(m)}) = m^{2}a_{m}r^{m-2} 
                                            \sim \epsilon^{m-2}$,
  $\frac{A^{(n)}}{r^{2}} = a_{n}r^{n-2}\sim \epsilon^{n-2}$,
  $[A^{3}]^{(p)} \sim \epsilon^{p}$ and
  $A^{(q)} = a_{q}r^{q}\sim \epsilon^{q}$,
%
%
where
%
$
   [A^{3}]^{(p)}=\sum_{i,j=1}^{p} A^{(i)}A^{(j)}A^{(p-i-j)}.
$
%
Again, by matching powers of $r$ we can derive a recursion
relationship for $a_{n}$. Setting $m=n$ and $p=q=n-2$ we obtain
\begin{equation}\label{hmm2}
   a_{n} = (f_{n-2}-a_{n-2})/(n^{2}-1)
\nonumber,
\end{equation}
where \mbox{$f_{p}=[A^{3}]^{(p)}/r^{p}$}. 

We should note that $a_{2n}=0$ for all $n$. Therefore, taking a power of r
out of our expansion for $A(r)$ we find,
\begin{equation}\label{squidge}
   \Psi = A(r)r^{i\theta} = rg(r)e^{i\theta}\nonumber,
\end{equation}
where \mbox{$g(r)=g(r^{2})= a_{1} + a_{3}r^{2}+ a_{5}r^{4}+ \cdots =
\sum_{n=1}^{\infty}a_{2n-1}r^{2n-2} $}. Further, \mbox{$re^{i\theta}$} is
complex so we can write \mbox{$re^{i\theta}=x+iy$} and hence our prototype 
solution, for a vortex pointing along the $z$-axis, is
   $\Psi = (x+iy)g(x^{2}+y^{2})$.
%
%

We can manipulate this prototype solution to get an initial condition for 
our vortex reconnection problem. Our initial condition $\Psi_0$ will be made 
up of two vortices, $\Psi_1$ and $\Psi_2$, a distance $2d$ and angle 
$2\alpha$ apart. Following the example of others, [Koplik {\em et al.}, Ref. \onlinecite{Koplik}] and
[Leadbeater {\em et al.}, Ref. \onlinecite{Leadbeater}], 
we take the initial condition to be the product of $\Psi_1$ and $\Psi_2$,
that is
   $\Psi_0=\Psi_1 \Psi_2\nonumber$.
%
%
One could argue that such an initial condition is rather special, as
two vortices found in close proximity would typically have already 
distorted one another
in their initial approach. Nevertheless, such a configuration provides us
with a valuable insight into the dynamics of NLS vortex reconnections.

Firstly, we would like the vortices in the $(x,y)$ plane. We can do this by
transforming our coordinates $x\rightarrow y$, $y\rightarrow z$
and $z\rightarrow x$. This will give us a vortex pointing along the $x$-axis 
   $\Psi = (y+iz)g(y^{2}+z^{2})\nonumber$.
%
%
The vortex can now be rotated by angle $\alpha$ to the $x$-axis in the 
$(x,y)$ plane via
      $x \rightarrow x\cos\alpha -y\sin\alpha$ and 
      $y \rightarrow y\cos\alpha +x\sin\alpha$.
%
%
Finally, we shift the whole vortex in the $z$ direction by a distance $d$ 
using \mbox{$z\rightarrow z-d$} we finally obtain
   $\Psi_{1} = [y\cos\alpha+x\sin\alpha+i(z-d)]
     g((y\cos\alpha+x\sin\alpha)^{2}+(z-d)^{2})$.
%
%
In a similar manner, $\Psi_2$ is a vortex at angle $-\alpha$ and shifted 
by $-d$ in the $z$ direction,
   $\Psi_{2} = [y\cos\alpha-x\sin\alpha+i(z+d)]
    g((y\cos\alpha-x\sin\alpha)^{2}+(z+d)^{2})$.
%
%

\section{RESULTS}
%
\begin{figure}[!Ht]
\centerline{\includegraphics[height=2in]{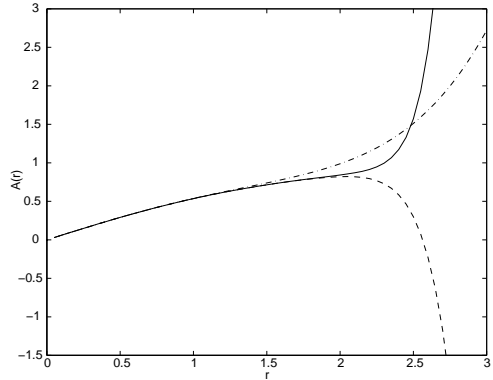}}
    \caption{The initial condition for the prototype vortex solution is constructed 
            via an appropriate expansion for \mbox{$A=A(r)$}, Eq. (\ref{hmm2}). 
            Here we can see the expansion for \mbox{$A=A(r)$} truncated at three 
            different orders of \mbox{$n$}; \mbox{$n=5$} (dash-dot line), \mbox{$n=15$} (dashed line) and 
            \mbox{$n=21$} (solid line) with \mbox{$a_1=0.6$}. At higher orders one would see the 
            existence of a finite radius of convergence at \mbox{$r\approx 2.5$}.} 
   \label{Ar}
\end{figure}
%
%
\begin{figure}[!Ht]
\centerline{\includegraphics[width=4.5in]{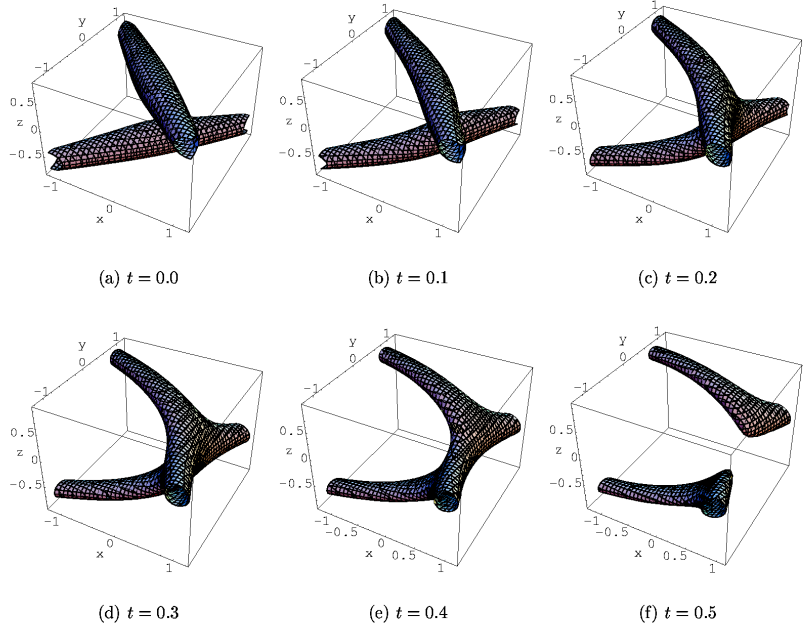}}
   \caption{Sub-figures (a) to (f) show the evolution of two initially separated vortices
            in time \mbox{$t$}. This realization is for \mbox{$|\Psi|=0.1$}, \mbox{$a_1=0.6$}, 
            \mbox{$d=0.6$} and \mbox{$\alpha=\pi/4$}. The reconnection and separation events 
            are clearly evident.}
   \label{pics}
\end{figure}
It would time consuming to expand the analytical solution, derived in the previous section, 
by hand. Thankfully, we can use a computer to perform the necessary algebra, and to derive
the hugh number of terms the recursive formulae generate. 
What follows is an example solution of the reconnection of two initially separated vortices.

Firstly, we need to consider the validity and accuracy of
our initial condition. Fig. \ref{Ar}, shows the prototype solution
$A=A(r)$ for a single vortex, at various different orders. 
Increasing the order will obviously improve accuracy. However, one should 
note that at higher order there is evidence of a finite radius of convergence
$r_c$. This will restrict the spatial region of validity for our full
t-dependent solution. Our prototype solution also has a dependence
on $a_1$. 
In the following simulation we have chosen $a_1=0.6$ numerically
so that the properties of $A(r)$ match that of a NLS vortex.
It is evident that we cannot satisfy these properties completely (namely
$\Psi\rightarrow 1$ as $r \rightarrow \infty$) as our power series 
diverges near $r_c$. Nevertheless, this does not present us with
a problem if we restrict ourselves to considering the evolution of 
contours of $\Psi$, such as $|\Psi|< 1$, where $A(r)$ is realistically represented. 
Further, it should be noted that sound radiation could in principle be visualized in 
our solution by drawing contours of $|\Psi|$ close to unity. However, to have 
an accurate representation, we would need to take a very large number of terms
in the series expansion, therefore the study of sound in our model is somewhat 
harder than the analysis of the vortices themselves. 
Of course the validity of the full $t$-dependent solution will be restricted,
in the spatial sense, by the initial condition's region of convergence. 
The region of convergence will evolve, remaining finite during a 
finite interval of time (by the Cauchy-Kowalevski theorem), but then may shrink to
zero afterwards. 

We will now discuss an example solution. 
As we only wish to demonstrate this method, we will not consider
a high order solution in this paper. In our example 
simulation below, we used Mathematica to perform the
necessary algebra in generating a nonlinear solution up to 
\mbox{$O(\epsilon^{6})$}. (One should note that although the 
prototype solution (\ref{squidge}) for a single vortex has $a_{2n}=0$ for all $n$, 
our initial condition is made up of two vortices, i.e. two series multiplied together. 
Therefore, there will be cross terms of order $O(\epsilon^{6})$ in our initial condition).


Our choice of parameters will be $d=0.6$ and $\alpha=\pi/4$. 
This corresponds to two vortices, initially separated  by a distance 1.2, 
at right angles to each other. Fig. \ref{pics} shows the evolution of the iso-surface 
$|\Psi|=0.1$ in time, demonstrating reconnection and then separation. Examining this 
solution in detail we can clearly see evidence of some of the properties 
mentioned earlier - that of a smooth reconnection (the absence of singularity)
and the anti-parallel alignment of vortices prior to reconnection.

\section{CONCLUSION}

In this paper we presented a local analysis of the NLS reconnection processes.
We showed that many interesting properties of the reconnection  can
already be seen at the linear level of the solution. We derived a recursion
formula Eq. (\ref{t-rec}) that gives the fully nonlinear solution of the 
initial value problem in a finite volume around the reconnection point for 
a finite period of time. In fact, formula (\ref{t-rec}) can describe a 
much wider class of problems. Of interest, for example, 
are solutions describing the creation or annihilation of 
NLS vortex rings. This process is easily described by considering vortex 
rings, at there creation/annihilation moment, as the
the intersection of a plane with the minimum of a paraboloid.
Further, this method of expansion around a reconnection point can be used 
for other evolution equations, e.g. the Ginzburg-Landau equation. 
These applications will be considered in future.
We wish to thank Robert Indik, Nicholas Ercolani and Yuri Lvov for  
their many fruitful discussions.







\end{document}